\documentstyle[manuscript,prb,aps]{revtex}
\include{epsf}

\begin{document}
\bibliographystyle{prsty}

\title{Analytical Rebridging Monte Carlo: 
Application to \emph{cis/trans} Isomerization in Proline-Containing, Cyclic Peptides
}
\author{Minghong G.\ Wu and Michael W.\ Deem \\
Chemical Engineering Department\\
University of California\\
Los Angeles, CA\ \ 90095-1592}
\maketitle
\begin{abstract}
We present a new method, the analytical rebridging scheme, for Monte
Carlo simulation of proline-containing, cyclic peptides. The {\em cis/trans}
isomerization is accommodated by allowing for two states of the amide
bond. We apply our method to five peptides that have previously been 
characterized by NMR methods.  Our simulations achieve effective
equilibration and agree well with experimental data in all cases. We
discuss the importance of effective equilibration as well
as the role of bond
flexibility and solvent effects in the
prediction of equilibrium properties.
\vspace{2in}
\end{abstract}
Submitted to \emph{J.\ Chem.\ Phys.}
\newpage
\section{Introduction}
One modern approach to therapeutics is to identify and to block the
molecular interactions responsible for disease. Within this context,
combinatorial peptide screening experiments play an important role in
the discovery of inhibitors. Libraries of peptides are used to
discover the crucial molecular structure, or pharmacophore, that can
block effectively the aberrant binding event. Cyclic peptides are
preferred for this purpose since they display higher binding affinities
due to their reduced conformational entropy in
solution.\cite{Jackson_94,Bisang_98} The cyclic scaffolds and
templates derived from combinatorial cyclic peptide studies have been
used to assemble various spatially-defined functional groups, and
highly active analogs have been synthesized in this
way.\cite{Bisang_98,Haubner_96,Huang_93} A classic example is the
blocking of the molecular event responsible for blood platelet
aggregation by peptides of the form
CRGDxxxC\put(-5,15){\line(-1,0){53}} \put(-5,10){\line(0,1){5}}
\put(-58,10){\line(0,1){5}}, CxxxRGDC\put(-5,15){\line(-1,0){53}}
\put(-5,10){\line(0,1){5}} \put(-58,10){\line(0,1){5}}, and
CxxxKGDC\put(-5,15){\line(-1,0){53}} \put(-5,10){\line(0,1){5}}
\put(-58,10){\line(0,1){5}}.\cite{II_ONeil} Organic analogs of these
RGD peptides have been synthesized, and several companies are pursuing
therapeutic applications in clinical trials.

The ability of a peptide molecule to bind selectively to a receptor
depends on its structural and conformational properties, which are in
turn determined by its constituent amino acids.  As the only natural
secondary amino acid, proline plays a particular role in peptide and
protein structural biology as a turn-promoting unit.\cite{Stryer_88}
Although most amide bonds in native peptides and proteins are in the
{\em trans} state, a significant fraction (6\%) of the X-Proline amide
bonds are in the {\em cis} state.\cite{Stryer_88} In some
proteins~\cite{Higgins_88} and cyclic peptides,\cite{Weisshoff_96}
prolyl {\em cis/trans} isomerization has been detected in solution.  This
isomerization event is frequently a rate-limiting step in protein
folding.  The free energy barrier for the {\em cis/trans} isomerization of
typical X-Pro amide bonds is about 19 Kcal/mol.\cite{Stein_93} While
isomerization of non-prolyl amide bonds is rare, the energy barrier to
isomerization is similar.\cite{Scherer_98}

Conventional molecular simulations cannot follow the natural {\em cis/trans}
isomerization dynamics, since the typical equilibration time is in the
range of 10 to 100 s.\cite{Schmid_93} Monte Carlo simulations have
been successful only in the case where isomerization is experimentally
known to occur. In such cases, an isomerization reaction coordinate
can be defined, and special techniques such as umbrella sampling can be
applied.  Conventional molecular dynamics for cyclic peptides, even in
the absence of {\em cis/trans} isomerization, turns out to be non-ergodic as
well, failing to sample the multiple solution conformations during the
accessible simulation time. The difficulties arise from the cyclic constraints
that, along with the intrinsic high energy barriers, isolate the
accessible conformations to several separated regions in phase
space. Monte Carlo methods, however, do not require the system
to follow the natural trajectory. Therefore, larger and unphysical
moves can be performed to overcome these energy barriers. A numerical
peptide rebridging method, inspired by the alkane
rebridging method~\cite{Dodd_93} and the configurational bias Monte Carlo
(CBMC) method,\cite{Frenkel_92,dePablo_92} has been successfully applied to the
simulation of non-proline-containing, cyclic peptides.\cite{Wu_99}
Parallel tempering was shown to be a key factor in the efficiency of
equilibration. Proline poses some geometrical complexity in the
rebridging approach, however, and has not yet been treated.

Here we present a new Monte Carlo method, the analytical rebridging
scheme, that is suitable for equilibration of proline-containing,
cyclic peptides. Our analytical method was inspired by the solution
for a related inverse kinematics problem in robotic
control.\cite{Manocha_94} The method can accommodate any rigid unit
geometry. The rebridging method is not restricted to peptides and can
be readily applied to other molecules.  The {\em cis/trans} isomerization is
naturally incorporated in the method by allowing for two states of the
amide bond. As an added benefit, we find the analytical approach to be
at least ten times more computationally efficient than the previous
numerical method,\cite{Wu_99} even in the simplest, non-proline-containing
cases. The analytical rebridging method and other
components of our simulation methodology are described in
Section~\ref{sec:methods}. We apply our method to five cyclic peptides
that have previously been characterized by NMR methods in
Section~\ref{sec:results}. We show that our method can effectively
equilibrate these molecules, yielding conformations consistent with
the NMR analyses. We discuss the results in
Section~\ref{sec:discussion} and conclude in
Section~\ref{sec:conclusion}.

\section{Simulation Methods}
\label{sec:methods}
\subsection{Analytical Peptide Rebridging}
\label{sec:methods-apr}
Our equilibration scheme involves three types of moves: rebridging
moves, semi-look-ahead (SLA) moves for side chains, and swapping
moves. The rebridging move is described below and in
Sec.~\ref{sec:methods-cti}. We describe the SLA move in
Sec.~\ref{sec:methods-sg} and the swapping move in
Sec.~\ref{sec:methods-pt}.

In our system, bond lengths and bond angles are kept at their
equilibrium value.  We focus on sampling the biologically-relevant,
torsional degrees of freedom. With this simplification, a molecule is
comprised of a set of rigid units.\cite{I_Deem1} The rebridging scheme
can easily be generalized to accommodate flexible bond angles and bond
lengths.

A peptide rebridging move causes a local conformational change within
the molecule, leaving the rest of the molecule fixed. Consider the
segment of a peptide backbone shown in Fig.~\ref{fig:reb_demo}. The
angles $\phi_{0}$ and $\phi_{7}$ are rotated, causing the rigid units
between 0 and 6 to change.  The range of rotation is within
$\pm\Delta\phi_{\mathrm{max}}$. These two rotations break the
connectivity of the molecule. We then find all the solutions that
re-insert the backbone units in a valid way between rigid units 1 to
6. The rebridging move is based upon enforcing
six geometrical constraints, and this is why we choose to modify six rigid
units.  Modification of more than six units in a single move is possible
by the rotation of more than two angles.
Such a move is likely, however, to lead to an infeasible
geometry most of the time and so to result in ineffective
equilibration. Therefore, we choose to rotate $\phi_{0}$ and $\phi_{7}$
only. 

Our peptide rebridging scheme features an analytical solution of
the geometrical problem arising from the reconnection of the backbone
units, a problem previously solved in a numerical
way.\cite{Wu_99,I_Deem1} The side chains are rigidly rotated for each of the
solutions. For rebridging moves, the solutions for both the new and
the old configurations are needed so as to satisfy detailed
balance. The analytical solution involves the reduction of twenty
linear equations to an 8$\times$8 determinant equation of one
torsional angle. The determinant equation is equivalent to a
polynomial of degree sixteen. Therefore, the maximum number of new
geometrical solutions is strictly limited to 16, a bound that is
obeyed in previous simulations using the numerical rebridging
method.~\cite{crankshaft} The determinant equation is reformulated as
an eigenvalue problem~\cite{Manocha_94} and solved using the QR
algorithm.\cite{C_recipes} The details of the analytical rebridging
method are described in Appendix~\ref{AppA}.

Following the ``with Jacobian'' (WJ) biasing,\cite{Wu_99} one of the
solutions is picked with a probability proportional to
${\mathrm J^{(n)}}\exp (-{\mathrm \beta U^{(n)}})$, the product
of the Jacobian and the Boltzmann factor of the solution.  The
Jacobian $\mathrm{J}$ accounts for the correction to the non-uniform
distribution generated by rebridging moves, and can be expressed in
several ways. Here we present one form, involving the
determinant of a $5\times 5$ matrix ${\mathbf B}$:
\begin{eqnarray}
\label{eqn:jac_RB}
{\mathrm J}
& = &    \frac{ {\hat{\mathbf u}}_{6}\cdot{\hat{\mathbf e}}_3 }
    { \det \vert{\mathrm B} \vert }\nonumber \\
{\mathbf B}_{ij} & = & 
			[{\hat{\mathbf u}}_{j}\times({\mathbf r}_{6}-{\mathbf r}_{\mathit{j}})]_i  \mbox{,
	if  $i\leq3$}, \nonumber  \\
& &                      [{\hat{\mathbf u}}_{j}\times{\hat{\mathbf u}}_{6}]_{i-3} \mbox{,  if $i=4,\
	5$} .
\end{eqnarray}
Here ${\hat{\mathbf u}}_{i}$ is the unit vector about which the torsional angle
$\phi_{i}$ is measured, ${\mathbf r}_{i}$ is the position of the atom in unit
$i-1$ that is bonded by a sigma bond to unit $i$, and ${\hat{\mathbf e}}_3$
 is a unit vector along the laboratory $z$-axis.  The subscript
outside the brackets refers to the component of the bracketed
vector in the laboratory frame. The Jacobian can also be written in
terms of a determinant of a $4\times 4$ matrix, although the
definition of the components of the $4\times 4$ matrix is more
involved.~\cite{Wu_99} The attempted move is accepted with the
probability
\begin{equation}
\label{eqn:acc-WJ}
	\mathrm{acc}(\mathrm{o\rightarrow n}) = \min \left(1,
	\frac{W^{\mathrm{(n)}}}{W^{\mathrm{(o)}}}\right)\ ,
\end{equation}
where ${\mathrm W^{(n)}}$ and ${\mathrm W^{(o)}}$ are the
normalization (Rosenbluth) factors for the new and old solutions,
respectively.\cite{Wu_99,Frenkel_96}

\subsection{The {\em cis/trans} Isomerization}
\label{sec:methods-cti}
For each amide bond that we wish to allow to isomerize, we assign two
discrete states to the corresponding rigid unit. As shown in
Fig.~\ref{fig:cistrans}, the amide unit takes the torsional values of
$\omega=0^\circ$ or $\omega=180^\circ$ in the {\em cis} and {\em
trans} conformations, respectively. The partition function includes a
summation over both states. Because of the sum over states, solutions
corresponding to all the possible {\em cis/trans} states between unit
0 and unit 6 are included in the calculation of the Rosenbluth
factor. The same approach can be applied to both non-prolyl and prolyl
amide bonds.
\subsection{Side Groups}
\label{sec:methods-sg}
The chemical functionality of peptides lies mostly in the side
chains. Efficient equilibration for the side chains, therefore, is
very important. The semi-look-ahead (SLA) move, based on CBMC methods,
has been shown to equilibrate effectively long and bulky
chains.\cite{Wu_99} We use the SLA method to equilibrate side chains
and end groups in this work. A SLA move proceeds by regrowing a
randomly selected side chain, unit by unit, beginning from the bond
that connects the backbone to the side chain. The reverse move is
performed so as to satisfy detailed balance. The Jacobian for each
solution is unity. The attempted move is accepted with the probability
\begin{equation}
\label{eqn:acc-SLA}
	\mathrm{acc}(\mathrm{o\rightarrow n}) = \min \left(1,
	\frac{W^{\mathrm{(n)}}}{W^{\mathrm{(o)}}}\right)\ ,
\end{equation}
where ${\mathrm W^{(n)}}$ and ${\mathrm W^{(o)}}$ are the
normalization (Rosenbluth) factors for the new and old geometries,
respectively.\cite{Frenkel_96}

\subsection{Parallel Tempering}
\label{sec:methods-pt}
Parallel tempering was first proposed for the study of glassy systems
with large free energy barriers.\cite{geyer_91} It has since been
successfully applied to a variety of
systems.\cite{hukushima_96,marinari_98,tesi_96,Boyd_98,Falcioni_99}
This method achieves rigorously correct canonical sampling, and it
significantly reduces the equilibration time in a simulation. Instead
of a single system, we consider in
parallel tempering a larger ensemble with $n$ systems,
each equilibrated at a distinct temperature $T_i$, $i=1,\ \ldots,\ n$. The
system with the lowest temperature is the one of our interest; the
higher temperature systems are added to aid in the equilibration of
the system of interest. In
addition to the normal Monte Carlo moves performed in each system,
swapping moves are proposed that exchange the configurations between
two systems $i$ and $j=i+1$, $1\leq i< n$.  A swapping move is
accepted with the probability
\begin{eqnarray}
\label{eqn:acc-pt}
{\mathrm acc}[(i,\ j)\rightarrow (j,\ i)] & = &
	\min[1,\exp(-\Delta\beta\Delta\mathrm{U})]\ ,
\end{eqnarray}
where $\Delta\beta$ and $\Delta\mathrm{U}$ are the difference of the
reciprocal temperatures and energies, respectively.  The higher
temperature systems are included solely to help the lowest temperature
system to escape from local energy minima via the swapping moves.  To
achieve efficient sampling, the highest temperature should be such
that no significant free energy barriers are observed. So that the
swapping moves are accepted with a reasonable probability, the energy
histograms of systems adjacent in the temperature ladder should
overlap.
\section{Comparison with Experimental Results}
\label{sec:results}
We perform simulations on five distinct cyclic peptides that have been
previously characterized by NMR methods.  We focus on the backbone
structure of proline-containing, cyclic peptides that were observed
experimentally to undergo {\em cis/trans}
isomerization. Molecular interactions are described by the AMBER force
field with explicit atoms.\cite{Weiner_86} Aqueous solvent effects are
estimated by simple dielectric theory.\cite{Smith_93} Five or six
systems were used in the parallel tempering for each simulation, with
the highest temperatures ranging from 10$^5$ K to 10$^7$ K.  The lowest
temperature system in each case is 298 K. For the first three
peptides, {\em cis/trans} isomerization is allowed in prolyl amide bonds
only. For the last two peptides, isomerization of all amide bonds is
allowed. The simulations take 5-8 CPU hours for the first three
peptides and 15-20 CPU hours for the last two peptides. All the
simulations were performed on an Intel Pentium II 450 MHz Linux
workstation. Rapid equilibration is achieved for all peptides.  The
results for each of the peptides are presented below:
\begin{enumerate}
\item c(Pro-Phe-D-Trp-Lys-Thr-Phe).  This analog of somatostatin
displays high activity in inhibiting the release of a growth
hormone.\cite{Huang_93,Veber_81} Analysis of the NMR spectrum in
D$_2$O solution indicated a unique backbone
conformation.\cite{Veber_81} The prolyl amide bond at Phe-Pro adopted a
{\em cis} conformation.  In our simulation, we find an essentially unique
conformation, possessing the same amide bond {\em cis/trans} sequence. A
representative conformation for this peptide is shown in
Fig.~\ref{fig:ex1}.

\item c(Phe-Phe-Aib-Leu-Pro).  This pentapeptide contains the
Pro-Phe-Phe sequence that has been proposed to be responsible for the
cytoprotective ability of antamanide and
cyclolinopeptides.\cite{Kessler_86,Zanotti_98} NMR analysis indicated that the
peptide is conformationally non-homogeneous at room
temperature. Two predominant {\em cis/trans} isomers for the Leu-Pro amide
bond were identified in acetonitrile at 240 K.\cite{Zanotti_98} Our
simulation led to two inter-converting conformers in the simulation,
as shown in Fig.~\ref{fig:ex2}. The {\em cis} and {\em trans} conformers occur
with probability 58\% and 42\%, respectively.


\item c(Gly-Glu(OBzl)-Pro-Phe-Leu-Pro). This cyclic hexapeptide was
synthesized for use as a possible chiral site for enantiomeric
separation.\cite{McEwen_93} NMR studies in dimethyl sulfoxide (DMSO)
reported two isomers, one having two {\em cis} prolyl bonds, and the other
having all-{\em trans} bonds.  We find only the 2-{\em cis} conformer in
simulation, as shown in Fig.~\ref{fig:ex4}.  All the torsional angles
fluctuate around mean values, except that the amide group between Pro
and Gly flips between two opposite orientations. The all-{\em trans}
conformer was found at higher temperatures.

\item c(Pro-Ala-Pro-Ala-Ala). This cyclic peptide
has been designed to serve as a rigid structural template. An unique
conformation with two {\em cis} prolyl amide bonds in DMSO solution was
found, according to NMR analysis.\cite{Mueller_93} The backbone
consists of two intertwined type-VIb $\beta$ turns,
centered about the two prolyl amide bonds respectively. We find
the same unique conformation. The torsional angles are close to the
values derived from restrained molecular
dynamics.\cite{Mueller_93} Figure~\ref{fig:ex7} depicts the geometry
of this peptide.
 
\item Tentoxin, c(MeAla-Leu-MePhe[(Z)$\triangle$]-Gly). This
tetrapeptide selectively induces chlorosis in the seedlings of
plants. Although tentoxin lacks proline, its two methylated amide
bonds were found to adopt the {\em cis} conformation in a nearly saturated
aqueous solution.\cite{Pinet_95} The other two non-methylated amide
bonds adopt the {\em trans} conformation.  The observation of this
{\em cis}-{\em trans}-{\em cis}-{\em trans} sequence of the backbone, along with other
experimental data, led to a proposed boat-like conformation, with the
two {\em cis} bonds located on the same side of the mean plane. In our
simulation, we find the same amide bond sequence and boat-like
conformation that was found in the experimental structure. The conformation
is shown in Fig.~\ref{fig:ex8}. All the carbonyl groups lie in the
same side of the mean plane, which implies that we have found the the
third major (conformer C, 8\% abundance) of the four conformers found at 268 K
in ref.~\onlinecite{Pinet_95}. The four conformers differ only in the
orientation of the two non-methylated amide groups.  The other three
conformers were found at higher temperatures in the simulation.
\end{enumerate}
\section{Discussion}
\label{sec:discussion}
By analyzing the energy trajectories and conformational data, we find
that all the peptides are effectively equilibrated with relatively few
Monte Carlo steps. For example, we find that the {\em cis/trans} equilibrium
for peptide 2 was attained within the first
10\% of the simulation time. Parallel tempering is crucial for this
inter-conversion, since the peptide essentially does not undergo 
\emph{cis/trans} isomerization in
a single, room-temperature, canonical simulation.

For peptide 1, the NMR-based conformational study indicated a
type-II'$\beta$ turn in the Phe-D-Trp-Lys-Thr region. This turn is
characterized by the hydrogen bond between the C=O of Phe and the N-H
of Thr. In our simulation, The C=O of Phe and the N-H of Thr are close
to each other, but are not in precise alignment. Such
disorder is expected at finite temperature.

Bond angle inflexibility may influence the predicted equilibrium
properties of these highly-strained molecules.  We have investigated
the dependence on bond angles by changing the angle between
C$_\alpha$-C and N-C$_\alpha$ in the amide bond on peptide 2 from
6$^\circ$ to 0$^\circ$. The predicted {\em cis/trans} equilibrium shifts from
42\% {\em trans} to 14\% {\em trans}, which is a non-trivial, although
energetically small, effect. We
suspect, therefore, that the absence of the all-{\em trans} conformer of
peptide 3 in our simulation may be due partly to inflexibility of the
bond angles.  We used a rigid proline ring, with a
$\phi_{\mathrm{Pro}}=-75^\circ$. This constraint suppresses the small
fluctuations that occur in the proline ring. In
exceptionally-constrained, cyclic peptides, $\phi_{\mathrm{Pro}}$ can take
on other values. For example, $\phi_{\mathrm{Pro}}\simeq -50^\circ$ has
been observed for some unusual prolines in the {\em trans} state.\cite{Mueller_93} 
Fluctuations in the proline ring may, therefore, be important in some
highly-strained systems.

Although experimental conformational analysis of peptide 3 in
water was not performed, NMR analysis in CHCl$_3$ indicated a minor
conformer in addition to the two major ones found in DMSO
solution.\cite{McEwen_93} It is clear, then, that solvent effects play
a role in the equilibrium structure of peptide 3. Note that the dielectric
constants are $\varepsilon_{\mathrm{CHCl_3}}=4.8$,\cite{CRC_87}
$\varepsilon_{\mathrm{DMSO}}=45$,\cite{Merck_83} and
$\varepsilon_{\mathrm{H_2O}}= 78$.\cite{CRC_87} The low-dielectric
CHCl$_3$ solution favored the all-{\em trans} conformer. Water, with a high
dielectric constant, may favor the 2-{\em cis} conformer, which is what we
observe in simulation. Indeed, upon reducing the dielectric constant
in our implicit solvent model, we find a small amount of the
all-{\em trans} state. However, the detailed structure of the solvent
molecules around the peptide is likely to be important, and a more
accurate description of solvent is likely necessary to account fully for the
solvent effects.

Tentoxin (peptide  5) in water  at  268 K  was experimentally found to
aggregate in a way  that suggested micellar organization. The critical
micelle  concentration  (CMC)   was   estimated   to  be   roughly  35
$\mu$M.\cite{Pinet_95} Conformations  observed at concentrations above
the CMC  may differ from that of  the dilute-limit, monomeric form. In
fact, of all four conformers found in ref.~\onlinecite{Pinet_95}, only
conformer  C  yielded chemical shifts   for the $\delta$  and $\gamma$
protons of  Leucine close to that  expected for  a monomeric form. The
other  two  major conformers,  A  and  B,  displayed strong  shielding
effects. The shielding constants  for the minor  conformer D  were not
reported.  These  shielding effects  were   explained by an aggregated
structure   of these  two  conformers.   Since  the concentration   of
conformer C  (250$\mu$M) is  still   well above  the  critical micelle
concentration, it is  not entirely clear why  the chemical shifts  for
conformer    C      were   relatively    unaffected    by    potential
aggregation.\cite{Pinet_95}   Nonetheless,     the   experimental data
suggests that conformer C  may be  either  in monomeric form or  in an
environment  similar to   aqueous  solution. This  gives  one possible
explanation for  the absence of conformers  A and B in our simulation,
since these conformers are certainly not in monomeric form.

\section{Conclusion}
\label{sec:conclusion}
We presented a new method, the analytical rebridging scheme, for the
simulation of chemically-diverse, chain-like molecules.  The method
naturally accommodates {\em cis/trans} isomerizations. Our rebridging scheme,
combined with parallel tempering and biased Monte Carlo, is very
successful at equilibration of proline-containing, cyclic
peptides. Our method is not limited to cyclic peptides and can
simulate any chain-like molecule. We compared our simulations with
experimental data on five cyclic peptides and found the predicted
conformations to be reasonably accurate.  We were able to sample
multiple, relevant conformations separated by high energy barriers, a
feat not possible with conventional molecular dynamics or Monte
Carlo. The numerical
quality of our predictions, while not limited by sampling issues, may
be limited by our choice of a simple forcefield. Nonetheless, our
method can be easily extended to accommodate flexible bond angles and
bond lengths. In addition, solvent effects may be represented more
accurately by better implicit solvent models.\cite{Ashbaugh_98,Paulaitis_97}

The methods described here are powerful enough and general enough to
influence the preferred approach to simulating biological systems. For
example, our method should be a valuable tool for the fitting of new
potential parameters for biological systems. New simulation methods
for long alkanes have made possible the optimization of force fields
that significantly reduce the discrepancies between simulation and
experiment.\cite{Smit_95,Nath_98} The same approach should lead to
improved forcefields for biological systems. In addition, we expect
that our peptide rebridging
scheme, combined with parallel tempering, should replace high temperature molecular dynamics and is readily
suitable for use in NMR-based conformational analyses of
biomolecules.

\section*{Acknowledgments}
This research was supported by the National Science Foundation through
grant no. CHE-9705165.

\newpage
\appendix
\section{Analytical Method for Solution of the Rebridging Problem}
\label{AppA}
Our analytical method was inspired by the inverse kinematics problem
of six-revolute manipulators, which is important for automatic control
of robotic arms.  This problem has been proved to have at most sixteen
solutions in the general case.\cite{Primrose_86} Lee and Liang reduced the problem to a
polynomial of one
variable of degree sixteen.\cite{LeeI_88,LeeII_88} The polynomial is derived by
equating the determinant of
an 8$\times$8 matrix to zero, each element being a quadratic
polynomial. A different closed form was obtained later by Raghavan and
Roth.\cite{Raghavan_93} An excellent review of this subject is given
by Manocha and Canny.\cite{Manocha_94} In the following, we
apply the symbolic formulation of Lee and Liang to reduce our peptide
rebridging problem to an eigenvalue problem.

The 6-revolute inverse kinematics problem can be formulated as an equivalent
closed, 7-revolute mechanism,\cite{LeeII_88} as shown in
Fig.~\ref{fig:7rlink}. The closed loop consists of seven joints with
offsets 
${\mathbf u}_1, {\mathbf u}_2,\ \ldots,\ {\mathbf u}_7$
and of seven links  with vectors
${\mathbf a}_1, {\mathbf a}_2,\ \ldots,\ {\mathbf a}_7$.
For a given backbone fragment to be
rebridged, the corresponding closed loop is uniquely determined by
calculating the link-joint intersections. We first draw the seven joint
lines that are parallel to the incoming sigma bonds of the seven rigid
units. The links are defined as the shortest vectors that
connect consecutive joint lines. Therefore, each link is perpendicular to
the two adjacent joint vectors. We denote the unit axes of the joint
${\mathbf u}_i$ and the link ${\mathbf a}_i$ by ${\hat{\mathbf u}}_{i}$
and ${\hat{\mathbf a}}_{i}$, respectively. The joint rotation
angles $\phi_{1},\ \ldots,\ \phi_{7}$ are the torsional angles
measured around the joints; they are equal to the corresponding
biological torsional angles plus constant offsets.  The joint rotation angle
$\phi_{7}$ and other parameters are determined by the given backbone
geometry, and we need to calculate only the six unknown joint rotation
angles.

The idea is to find an over-constrained set of twenty equations that
are linear in $x_6\equiv \tan\phi_{6}/2$ and the sines of cosines of
$\phi_{1},\ \phi_{2},\ \phi_{4}$, and $\phi_{5}$.  These equations are
obtained by equating scalar and vector products of the loop axes in
both directions of the loop. For instance, we equate
${\hat{\mathbf u}}_{3}\cdot{\hat{\mathbf u}}_{6}(\phi_{4},\ \phi_{5})={\hat{\mathbf u}}_{6}\cdot{\hat{\mathbf u}}_{3}(\phi_{1},\
\phi_{2})$ in the first equation (see below).  These equations were
first derived using a recursive notation.\cite{LeeII_88} We define the
chain vector ${\mathbf R}^{\alpha,\ \beta}$ as the vector summation of the
consecutive joints and links from unit $\alpha$ to unit $\beta$. The
summation always goes in the direction of units $1, 2,\ \ldots,\ 7$. The
indices $\alpha$ and $\beta$ can take one of two forms: i denotes
either starting from or ending at ${\mathbf u}_i$, and i' denotes
either starting from or ending at ${\mathbf a}_i$. The index $\beta$
can be less than $\alpha$, and this indicates wrapping around the
closed loop. For example,
\begin{eqnarray}
{\mathbf R}^{6',\ 2} &= & 
{\mathbf a}_6 + {\mathbf u}_7 +
{\mathbf a}_7 + {\mathbf u}_1 +
{\mathbf a}_1 + {\mathbf u}_2
\nonumber \\
{\mathbf R}^{i',\ i} &= & {\mathbf 0} \ .
\end{eqnarray}
The twenty equations are listed below. The left-hand side of each
equation is a linear function of $\cos\phi_{1},\ \sin\phi_{1},\ \cos\phi_{2},\
\sin\phi_{2},\ $ and $x_6$. The right-hand side of each equation is a
linear function of $\cos\phi_{4},\ \sin\phi_{4},\ \cos\phi_{5},\
\sin\phi_{5},\ $ and $x_6$:
\begin{eqnarray*}
{\hat{\mathbf u}}_{3}\cdot{\hat{\mathbf u}}_{6}	&=&{\hat{\mathbf u}}_{6}\cdot{\hat{\mathbf u}}_{3} \\
{\hat {\mathbf u}}_{3}\cdot{\hat{\mathbf u}}_{6}\ x_6	&=&{\hat{\mathbf u}}_{6}\cdot{\hat{\mathbf u}}_{3}\ x_6 \\
{\mathbf R}^{3',\ 5'}\cdot{\hat{\mathbf u}}_{6}\times{\hat{\mathbf u}}_{3}&=&{\mathbf R}^{6',\ 2'}\cdot{\hat{\mathbf u}}_{3}\times{\hat{\mathbf u}}_{6} \\
{\mathbf R}^{3',\ 5'}\cdot{\hat{\mathbf u}}_{6}\times{\hat{\mathbf u}}_{3}\ x_6&=&
		{\mathbf R}^{6',\ 2'}\cdot{\hat{\mathbf u}}_{3}\times{\hat{\mathbf u}}_{6}\ x_6 \\
{\mathbf R}^{3,\ 5'}\cdot{\hat{\mathbf u}}_{3}		&=&-{\mathbf R}^{6,\ 2'}\cdot{\hat{\mathbf u}}_{3} \\
{\mathbf R}^{3,\ 5'}\cdot{\hat{\mathbf u}}_{3}\ x_6	&=&-{\mathbf R}^{6,\ 2'}\cdot{\hat{\mathbf u}}_{3}\ x_6 \\
{\mathbf R}^{3,\ 5'}\cdot{\hat{\mathbf u}}_{6}		&=&-{\mathbf R}^{6,\ 2'}\cdot{\hat{\mathbf u}}_{6} \\
{\mathbf R}^{3,\ 5'}\cdot{\hat{\mathbf u}}_{6}\ x_6	&=&-{\mathbf R}^{6,\ 2'}\cdot{\hat{\mathbf u}}_{6}\ x_6 \\
{\mathbf R}^{3,\ 5'}\cdot{\mathbf R}^{3,\ 5'}	&=&{\mathbf R}^{6,\ 2'}\cdot{\mathbf R}^{6,\ 2'}\\
{\mathbf R}^{3,\ 5'}\cdot{\mathbf R}^{3,\ 5'}\ x_6 
				&=&{\mathbf R}^{6,\ 2'}\cdot{\mathbf R}^{6,\ 2'}\ x_6
\\
{\hat{\mathbf u}}_{3}\cdot{\hat{\mathbf a}}_{5}\ x_6 - {\hat{\mathbf u}}_{3}\cdot{\hat{\mathbf u}}_{6}\times{\hat{\mathbf a}}_{5}
			&= &-{\hat{\mathbf a}}_{6}\cdot{\hat{\mathbf u}}_{3}\ x_6 -
			{\hat{\mathbf u}}_{6}\times{\hat{\mathbf a}}_{6}\cdot{\hat{\mathbf u}}_{3} 
\\
{\hat{\mathbf u}}_{3}\cdot{\hat{\mathbf u}}_{6}\times{\hat{\mathbf a}}_{5}\ x_6 +{\hat{\mathbf u}}_{3}\cdot{\hat{\mathbf a}}_{5} 
			&= &-{\hat{\mathbf u}}_{6}\times{\hat{\mathbf a}}_{6}\cdot{\hat{\mathbf u}}_{3}\ x_6 +
			{\hat{\mathbf a}}_{6}\cdot{\hat{\mathbf u}}_{3} \\
-{\mathbf R}^{3,\ 5'}\cdot{\hat{\mathbf a}}_{5}\ x_6 + {\mathbf R}^{3,\ 5'}\cdot{\hat{\mathbf u}}_{6}\times{\hat{\mathbf a}}_{5}	
		&=&-{\mathbf R}^{6,\ 2'}\cdot{\hat{\mathbf a}}_{6}\ x_6 - {\mathbf R}^{6,\ 2'}\cdot{\hat{\mathbf u}}_{6}\times{\hat{\mathbf a}}_{6}
\\	
{\mathbf R}^{3,\ 5'}\cdot{\hat{\mathbf u}}_{6}\times{\hat{\mathbf a}}_{5}\ x_6 + {\mathbf R}^{3,\ 5'}\cdot{\hat{\mathbf a}}_{5}
		&=&{\mathbf R}^{6,\ 2'}\cdot{\hat{\mathbf u}}_{6}\times{\hat{\mathbf a}}_{6}\ x_6 - {\mathbf R}^{6,\
		2'}\cdot{\hat{\mathbf a}}_{6}
\end{eqnarray*}
\begin{eqnarray*}
&&{\mathbf R}^{3',\ 5}\cdot{\hat{\mathbf a}}_{5}\times{\hat{\mathbf u}}_{3}\ x_6 
	- {\mathbf R}^{3',\ 5'}\cdot({\hat{\mathbf u}}_{6}\times{\hat{\mathbf a}}_{5})\times{\hat{\mathbf u}}_{3}
\\
&&\;\;\;\;\;=-({\mathbf R}^{7,\ 2'}\cdot{\hat{\mathbf u}}_{3}\times{\hat{\mathbf a}}_{6}
				-{\mathbf{u}}_{6}\times{\hat{\mathbf a}}_{6}\cdot{\hat{\mathbf u}}_{3})\ x_6
			   -({\mathbf R}^{6',\ 2'}\cdot{\hat{\mathbf u}}_{3}\times({\hat{\mathbf u}}_{6}\times{\hat{\mathbf a}}_{6})
			+\vert {\mathbf u}_6 \vert {\hat{\mathbf a}}_{6}\cdot{\hat{\mathbf u}}_{3})
\\
&&{\mathbf R}^{3',\ 5'}\cdot({\hat{\mathbf u}}_{6}\times{\hat{\mathbf a}}_{5})\times{\hat{\mathbf u}}_{3}\ x_6 + {\mathbf R}^{3',\
5}\cdot{\hat{\mathbf a}}_{5}\times{\hat{\mathbf u}}_{3} 
\\ 
&&\;\;\;\;\; =-({\mathbf R}^{6',\
2'}\cdot{\hat{\mathbf u}}_{3}\times({\hat{\mathbf u}}_{6}\times{\hat{\mathbf a}}_{6}) +\vert {\mathbf u}_6 \vert {\hat{\mathbf a}}_{6}\cdot{\hat{\mathbf u}}_{3})\ x_6
+({\mathbf R}^{7,\ 2'}\cdot{\hat{\mathbf u}}_{3}\times{\hat{\mathbf a}}_{6}
-{\mathbf{u}}_{6}\times{\hat{\mathbf a}}_{6}\cdot{\hat{\mathbf u}}_{3}) 
\\
&&1/2({\mathbf R}^{3,\ 5'}\cdot{\mathbf R}^{3,\ 5'})({\hat{\mathbf u}}_{3}\cdot{\hat{\mathbf u}}_{6})
	-({\mathbf R}^{3,\ 5'}\cdot{\hat{\mathbf u}}_{3})({\mathbf R}^{3,\ 5'}\cdot{\hat{\mathbf u}}_{6})
\\
&&\;\;\;\;\;	=1/2({\mathbf R}^{6,\ 2'}\cdot{\mathbf R}^{6,\ 2'})({\hat{\mathbf u}}_{6}\cdot{\hat{\mathbf u}}_{3})
	-({\mathbf R}^{6,\ 2'}\cdot{\hat{\mathbf u}}_{6})({\mathbf R}^{6,\ 2'}\cdot{\hat{\mathbf u}}_{3}) \\
&&\left[1/2({\mathbf R}^{3,\ 5'}\cdot{\mathbf R}^{3,\ 5'})({\hat{\mathbf u}}_{3}\cdot{\hat{\mathbf u}}_{6})
	-({\mathbf R}^{3,\ 5'}\cdot{\hat{\mathbf u}}_{3})({\mathbf R}^{3,\ 5'}\cdot{\hat{\mathbf u}}_{6})\right]x_6
\\
&&\;\;\;\;\;    =\left[1/2({\mathbf R}^{6,\ 2'}\cdot{\mathbf R}^{6,\ 2'})({\hat{\mathbf u}}_{6}\cdot{\hat{\mathbf u}}_{3})
	-({\mathbf R}^{6,\ 2'}\cdot{\hat{\mathbf u}}_{6})({\mathbf R}^{6,\
2'}\cdot{\hat{\mathbf u}}_{3})\right]x_6 
\end{eqnarray*}
\begin{eqnarray}
\label{eqn:twenty_eq}
&&\left[1/2({\mathbf R}^{3,\ 5'}\cdot{\mathbf R}^{3,\ 5'})({\hat{\mathbf u}}_{3}\cdot{\hat{\mathbf a}}_{5})
	-({\mathbf R}^{3,\ 5'}\cdot{\hat{\mathbf a}}_{5})({\mathbf R}^{3,\
		5'}\cdot{\hat{\mathbf u}}_{3})\right]x_6 \nonumber
\\
&&\;\;\;\;\;\;\;\;\;\;\;\;\;\; -\left[1/2({\mathbf R}^{3,\ 5'}\cdot{\mathbf R}^{3,\ 5'})
				({\hat{\mathbf u}}_{3}\cdot{\hat{\mathbf u}}_{6}\times{\hat{\mathbf a}}_{5})
		-({\mathbf R}^{3,\ 5'}\cdot{\hat{\mathbf u}}_{3})
				({\mathbf R}^{3,\ 5'}\cdot{\hat{\mathbf u}}_{6}\times{\hat{\mathbf a}}_{5})\right]\nonumber
\\ 
&&\;\;\;\;\;	=-\left[1/2({\mathbf R}^{6,\ 2'}\cdot{\mathbf R}^{6,\ 2'})
				({\hat{\mathbf u}}_{3}\cdot{\hat{\mathbf a}}_{6})
		-({\mathbf R}^{6,\ 2'}\cdot{\hat{\mathbf a}}_{6})({\mathbf R}^{6,\
		2'}\cdot{\hat{\mathbf u}}_{3})\right]x_6 \nonumber
\\
&&\;\;\;\;\;\;\;\;\;\;\;\;\;\;\;\;\;\;\;\;\;\;\;\;\;\;	-\left[1/2({\mathbf R}^{6,\ 2'}\cdot{\mathbf R}^{6,\ 2'})
				({\hat{\mathbf u}}_{3}\cdot{\hat{\mathbf u}}_{6}\times{\hat{\mathbf a}}_{6})
		-({\mathbf R}^{6,\ 2'}\cdot{\hat{\mathbf u}}_{3})
			({\mathbf R}^{6,\ 2'}\cdot{\hat{\mathbf u}}_{6}\times{\hat{\mathbf a}}_{6})\right]\nonumber
\\
&&\left[1/2({\mathbf R}^{3,\ 5'}\cdot{\mathbf R}^{3,\ 5'})
				({\hat{\mathbf u}}_{3}\cdot{\hat{\mathbf u}}_{6}\times{\hat{\mathbf a}}_{5})
		-({\mathbf R}^{3,\ 5'}\cdot{\hat{\mathbf u}}_{3})
				({\mathbf R}^{3,\
		5'}\cdot{\hat{\mathbf u}}_{6}\times{\hat{\mathbf a}}_{5})\right]x_6\nonumber 
\\
&&\;\;\;\;\;\;\;\;\;\;\;\;\;\;	+\left[1/2({\mathbf R}^{3,\ 5'}\cdot{\mathbf R}^{3,\ 5'})
			({\hat{\mathbf u}}_{3}\cdot{\hat{\mathbf a}}_{5})
	-({\mathbf R}^{3,\ 5'}\cdot{\hat{\mathbf a}}_{5})({\mathbf R}^{3,\ 5'}\cdot{\hat{\mathbf u}}_{3})\right]\nonumber
 \\
&&\;\;\;\;\;	= -\left[1/2({\mathbf R}^{6,\ 2'}\cdot{\mathbf R}^{6,\ 2'})
				({\hat{\mathbf u}}_{3}\cdot{\hat{\mathbf u}}_{6}\times{\hat{\mathbf a}}_{6})
		-({\mathbf R}^{6,\ 2'}\cdot{\hat{\mathbf u}}_{3})
			({\mathbf R}^{6,\
		2'}\cdot{\hat{\mathbf u}}_{6}\times{\hat{\mathbf a}}_{6})\right]x_6\nonumber
\\
&&\;\;\;\;\;\;\;\;\;\;\;\;\;\;\;\;\;\;\;\;\;\;\;\;\;\;	+\left[1/2({\mathbf R}^{6,\ 2'}\cdot{\mathbf R}^{6,\ 2'})	
			({\hat{\mathbf u}}_{3}\cdot{\hat{\mathbf a}}_{6})
	-({\mathbf R}^{6,\ 2'}\cdot{\hat{\mathbf a}}_{6})({\mathbf R}^{6,\ 2'}\cdot{\hat{\mathbf u}}_{3})\right]\; .
\end{eqnarray}
This set of equations can be put into matrix form as
\begin{equation}
\label{eqn:20eq_mtx}
{\mathbf AB}= {\mathbf P}x_6 + {\mathbf Q}\ ,
\end{equation}
where ${\mathbf A}$ is a $20\times 16$ constant matrix, and ${\mathbf B}$ is a
$16\times 1$ matrix with the following variables as its elements:
\begin{eqnarray}
{\mathbf B}^\top=[c_4c_5x_6,\ s_4c_5x_6,\ c_5x_6,\ c_4s_5x_6,\
s_4s_5x_6,\ s_5x_6,\ c_4x_6,\ s_4x_6,\nonumber \\
c_4c_5,\ s_4c_5,\ c_5,\ c_4s_5,\
s_4s_5,\ s_5,\ c_4,\ s_4]\ .
\end{eqnarray}
Here $c_i=\cos\phi_{i}$ and $s_i=\sin\phi_{i}$. The $16\times 1$ matrices
${\mathbf P}$ and ${\mathbf Q}$ have elements that are linear in the sines and
cosines of $\phi_{1}$ and $\phi_{2}$. Sixteen equations are chosen from the
twenty equations in eq.~(\ref{eqn:twenty_eq}) and used to express ${\mathbf
B}$ as functions of $\phi_{1}$, $\phi_{2}$, and $x_6$. This is
accomplished by defining ${\mathbf A}_{\mathrm s}$ as the corresponding
$16\times 16$ sub-matrix of ${\mathbf A}$ and multiplying both sides of
these sixteen equations by the inverse of ${\mathrm A}_{\mathrm s}$ to obtain
\begin{equation}
\label{eqn:16x16_mtx}
{\mathbf B}= {{\mathbf A}_{\mathrm s}}^{-1}{\mathbf P}_{\mathrm s} x_6 + 
			{{\mathbf A}_{\mathrm s}}^{-1}{\mathbf
			Q}_{\mathrm s} \ .
\end{equation}
Here ${\mathbf P}_{\mathrm s}$ and ${\mathbf Q}_{\mathrm s}$ are the
corresponding sub-matrices of ${\mathbf P}$ and ${\mathbf Q}$.  We find that
special geometries, such as vanishing lengths of links, may render
some choices for ${\mathbf A}_{\mathrm s}$ singular.  The linear
dependence is identified by singular value decomposition
\cite{C_recipes} and avoided by choosing 16 linearly-independent
equations. The expression for ${\mathbf B}$ is inserted into the other four
unused equations to yield four equations that are linear functions of
$x_6$ and the sines and cosines of $\phi_{1}$ and $\phi_{2}$. Replacing
the sines and cosines of $\phi_{1}$ and $\phi_{2}$ with $x_1\equiv
\tan\phi_{1}/2$ and $x_2\equiv \tan\phi_{2}/2$, these four equations can
be expressed as
\begin{equation}
\label{eqn:four_eq}
(d_i{x_2}^2+e_ix_2+f_i)x_6+
	(g_i{x_2}^2+h_ix_2+p_i)=0,\;\;\;  i=1,\ \ldots,\ 4,
\end{equation}
where $d_i,\ e_i,\ \ldots,\ p_i$ are quadratic in
$x_1$. Multiplying eq.~(\ref{eqn:four_eq}) by $x_2$ gives four additional
equations, which together with eq.~(\ref{eqn:four_eq}) can
be used to set up a linear set of equations:
\begin{equation}
\label{eqn:8x8_mtx}
\left[ \begin{array}{cccccccc}
	0  &0  &d_1&e_1&f_1&g_1&h_1&p_1 \\
	0  &0  &d_2&e_2&f_2&g_2&h_2&p_2 \\
	0  &0  &d_3&e_3&f_3&g_3&h_3&p_3 \\
	0  &0  &d_4&e_4&f_4&g_4&h_4&p_4 \\
	d_1&g_1&e_1&f_1&0  &h_1&p_1&0   \\
	d_2&g_2&e_2&f_2&0  &h_2&p_2&0   \\
	d_3&g_3&e_3&f_3&0  &h_3&p_3&0   \\
	d_4&g_4&e_4&f_4&0  &h_4&p_4&0 \end{array} \right]
\left[	\begin{array}{c}
	{x_2}^3x_6 \\
	{x_2}^3	   \\	
	{x_2}^2x_6 \\
	 x_2 x_6   \\
	 x_6       \\
	{x_2}^2    \\
	 x_2       \\
	 1          \end{array}\right] = \mathbf{0}\; .
\end{equation}
Here ${\mathbf 0}$ is an $8\times 1$ null vector. Solutions are found by
equating to zero the corresponding determinant of the $8\times 8$
matrix, which can be expanded to a polynomial of degree sixteen in $x_1$.
The determinant equation is best solved
by reformulating it as an eigenvalue problem. This is achieved by
rewriting
eq.~(\ref{eqn:8x8_mtx}) as
\begin{equation}
\left|{\mathbf A}_0 + {\mathbf A}_1 x_1 
				+ {\mathbf A}_2{x_1}^2\right|=0\ ,
\end{equation}
where ${\mathbf A}_0$, ${\mathbf A}_1$, and ${\mathbf A}_2$ are
$8\times 8$ numerical matrices. Note that in the crankshaft case,
${\mathbf A}_0$, ${\mathbf A}_1$, and ${\mathbf A}_2$ are identically
zero, and so any solution for $x_1$ is possible.\cite{crankshaft} The
roots of the determinant equation are the eigenvalues of the
matrix~\cite{Manocha_94}
\begin{equation}
\left[ \begin{array}{cc} \mathbf{0} & \mathbf{I} \\
	-{{\mathbf A}_2}^{-1}{\mathbf A}_0 & -{{\mathbf
A}_2}^{-1}{\mathbf A}_1
	\end{array}\right]\ ,
\end{equation}
where $\mathbf{0}$ and $\mathbf{I}$ are $8\times 8$ null and identity
matrices, respectively. The matrix is first reduced to an upper
Hessenberg matrix, and then the eigenvalues are
found by the QR algorithm~\cite{C_recipes}
to obtain $x_1$. Occasionally the matrix ${\mathbf A}_2$ may be almost
singular. This occurs when one of the roots is
$\phi_{1}\approx \pm \pi$. This singularity can be avoided by the
transformation
\begin{equation}
	x_1 = \frac{t_1 \overline{x}_1 + t_2}{t_3\overline{x}_1 +
	t_4}\ ,
\end{equation}
where $t_1,\ t_2,\ t_3$, and $t_4$ are random numbers uniformly
distributed in [-1,1].  Solutions for $x_1$ are substituted into
eq.~(\ref{eqn:8x8_mtx}) to calculate $x_2$ and $x_6$. These values are
substituted back to eq.~(\ref{eqn:16x16_mtx}) to calculate $\phi_{4}$
and $\phi_{5}$. The angle $\phi_{3}$ is not needed for our purpose; it
can be determined by calculating the positions of ${\hat{\mathbf a}}_{2}$ and
${\hat{\mathbf a}}_{3}$ from the other torsional angles.  

\bibliography{proline}

\begin{thebibliography}{10}
\bibitem{crankshaft}
The analytical rebridging equations are singular for the geometry which allows
  for an infinite number of ``crankshaft-type'' solutions. This geometry occurs
  with probability zero in the general case.
\end{thebibliography}
\pagestyle{empty}
\begin{flushleft}
\begin{figure}[p]
\caption{A backbone segment selected to be rebridged. Only the
backbone atoms are shown. A change of the driver angles $\phi_{0}$ and $\phi_{7}$
breaks the connectivity. The dotted area represents the region in
which the positions of the backbone atoms must be
restored. The thick solid lines represent pi bonds or rigid
molecular fragments within which no rotation is possible.}
\label{fig:reb_demo}
\end{figure}
\begin{figure}[p]
\caption{The (a){\em cis} and (b){\em trans} conformation of the amide
bond. Only backbone atoms are depicted. The R$_{\mathrm{x}}$ atom is
hydrogen for a normal, non-prolyl amide bond or carbon
for a prolyl or methylated amide bond.}
\label{fig:cistrans}
\end{figure}
\begin{figure}[p]
\caption{The conformation of c(Pro-Phe-D-Trp-Lys-Thr-Phe). The colors
green, red, and blue denote carbon, oxygen,
and nitrogen atoms, respectively. The hydrogen atoms are omitted.}
\label{fig:ex1}
\end{figure}
\begin{figure}[p]
\caption{The (a){\em cis} and (b){\em trans} conformations of
c(Phe-Phe-Aib-Leu-Pro).\hspace{3in}}
\label{fig:ex2}
\end{figure}
\begin{figure}[p]
\caption{The conformation of c(Gly-Glu(OBzl)-Pro-Phe-Leu-Pro).\hspace{3in}}
\label{fig:ex4}
\end{figure}
\begin{figure}[p]
\caption{The conformation of c(Pro-Ala-Pro-Ala-Ala).\hspace{3in}}
\label{fig:ex7}
\end{figure}
\begin{figure}[p]
\caption{The conformation of
c(MeAla-Leu-MePhe[(Z)$\triangle$]-Gly).\hspace{3in}}
\label{fig:ex8}
\end{figure}
\begin{figure}[p]
\caption{The geometry of the closed, 7-revolute mechanism,
consisting of 7 joints and 7 links. The joints are represents by
${\mathbf u}_1,\ {\mathbf u}_2,\ \ldots,\ {\mathbf u}_7$.
 The links are represented by ${\mathbf a}_1,\
{\mathbf a}_2,\ \ldots,\ {\mathbf a}_7$.
 Each link is perpendicular in three dimensions to the two adjacent
joints. The unit axes of the joints and links are defined as
$\hat{\mathbf u}_{i}$ and $\hat{\mathbf a}_{i}$, respectively.}
\label{fig:7rlink}
\end{figure}
\end{flushleft}

\clearpage
\newpage
\begin{center}
\epsfxsize=6in \centerline{ \epsfbox{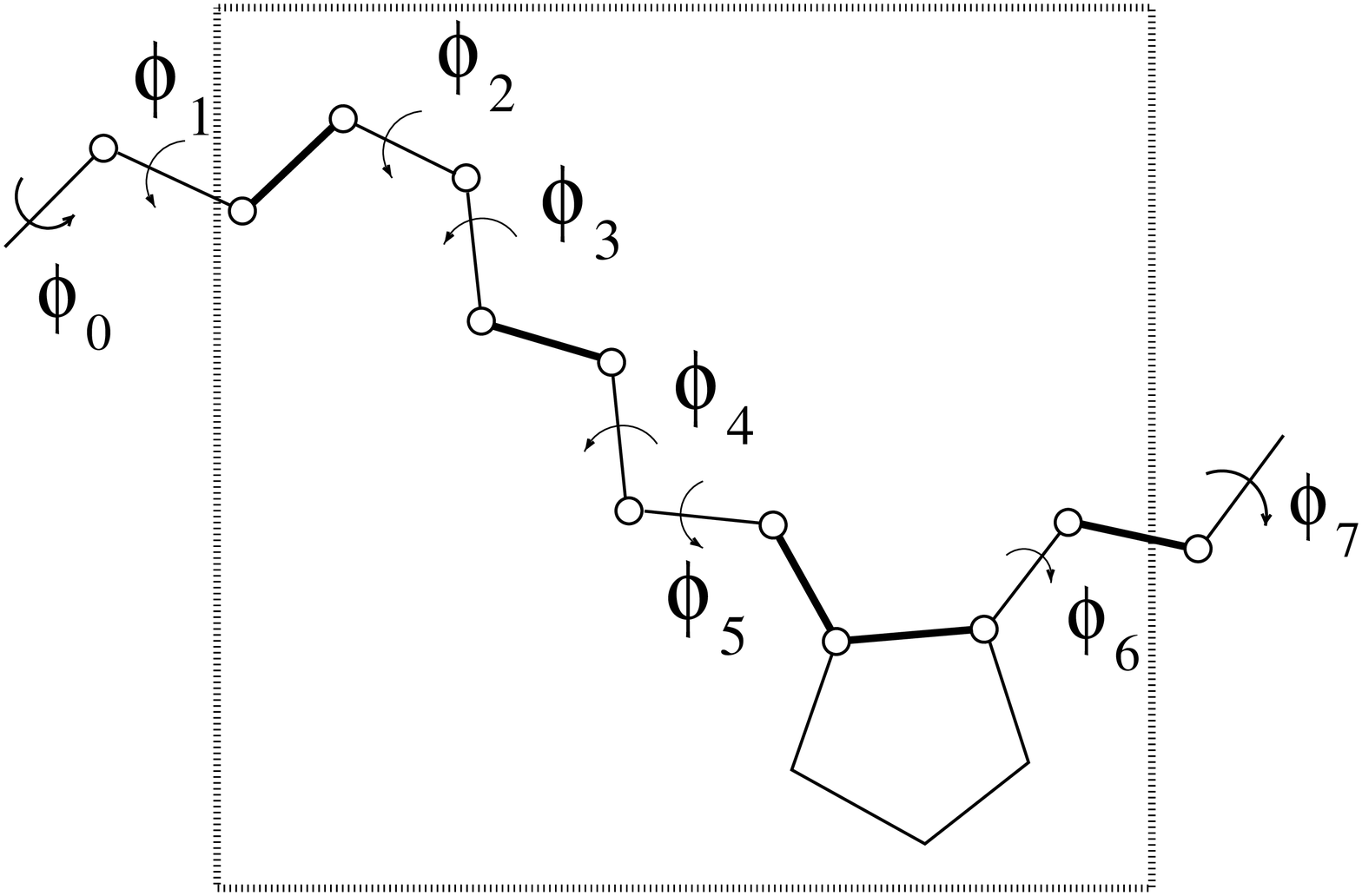}}
\vspace{1.5in}
Figure \ref{fig:reb_demo}.
 Wu and Deem, ``Analytical Rebridging Monte Carlo\ldots.''

\newpage
\epsfxsize=4in \centerline{ \epsfbox{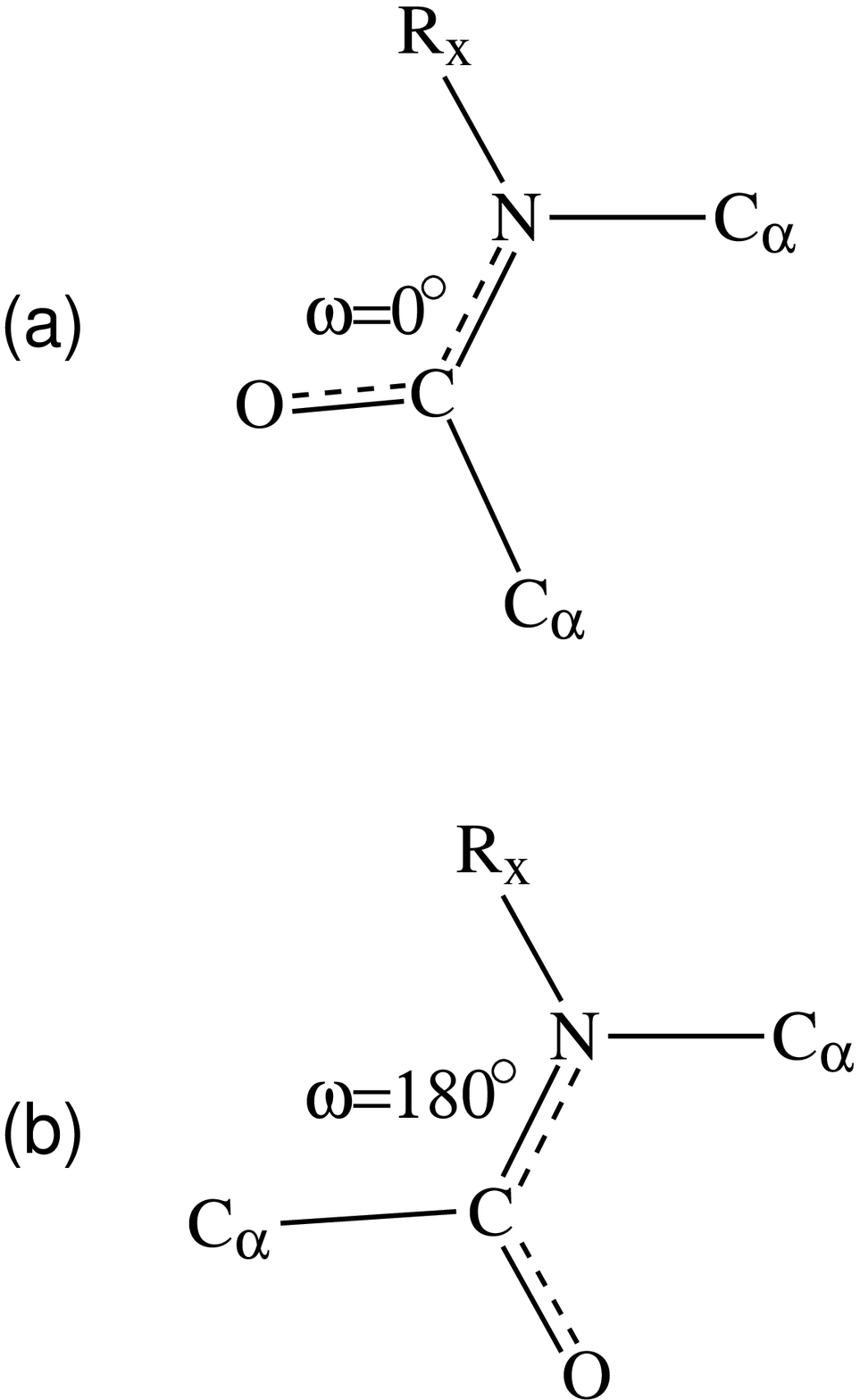}}
\vspace{1.5in}
Figure \ref{fig:cistrans}.
 Wu and Deem, ``Analytical Rebridging Monte Carlo\ldots.''

\newpage
\epsfxsize=5in \centerline{ \epsfbox{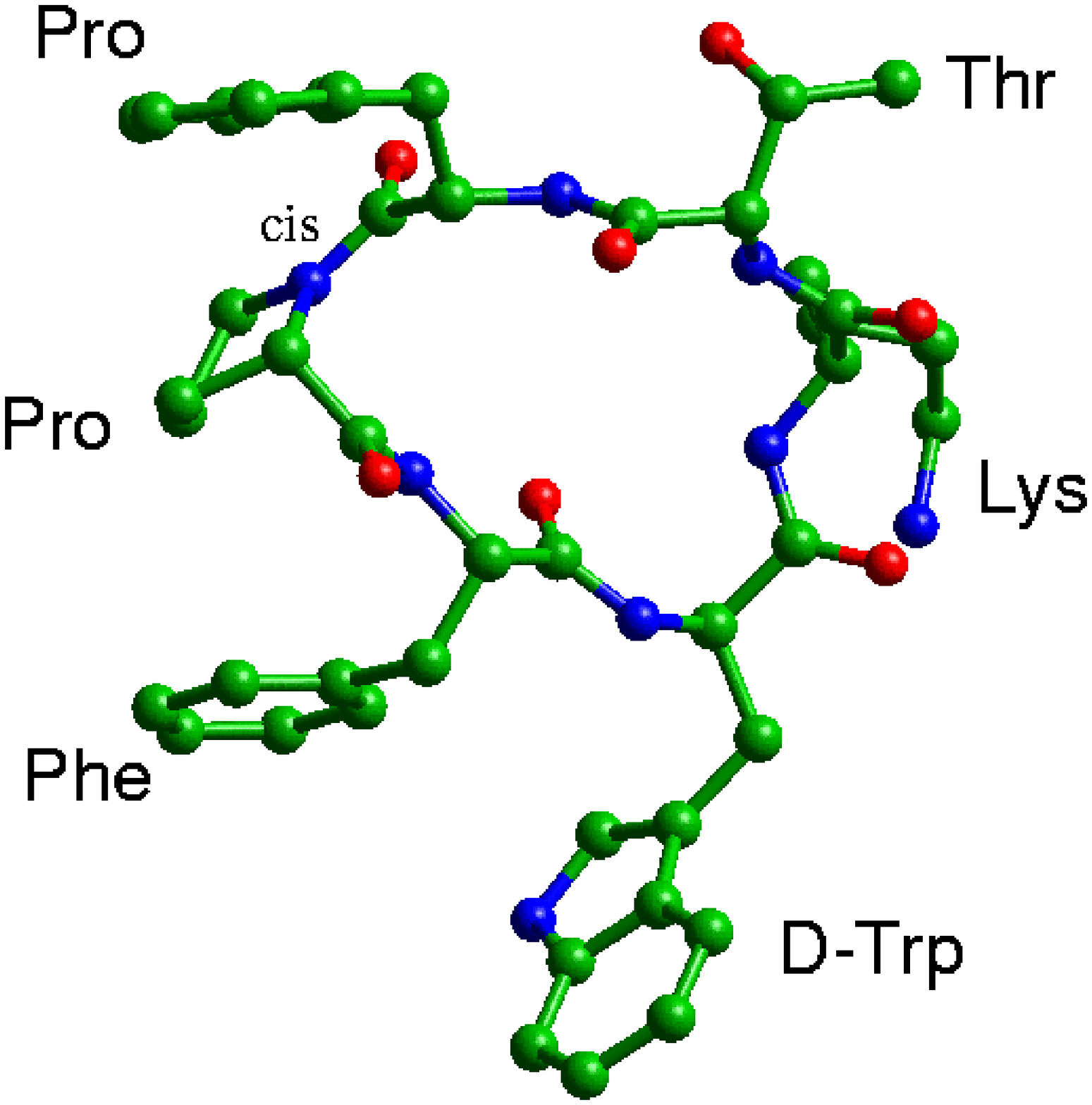}}
\vspace*{0.5in}
Figure \ref{fig:ex1}.
 Wu and Deem, ``Analytical Rebridging Monte Carlo\ldots.''

\newpage
\epsfxsize=4in \centerline{ \epsfbox{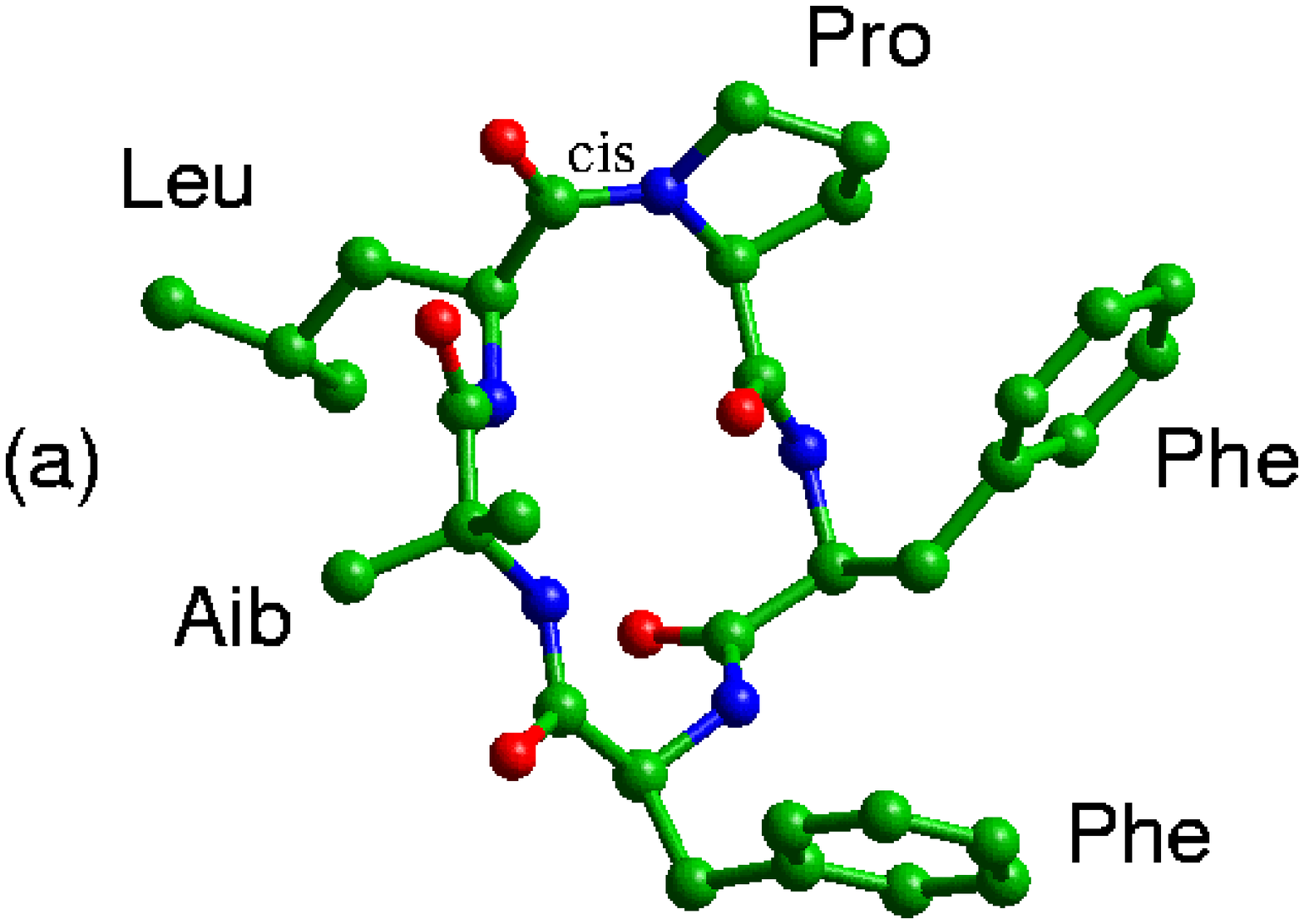}}
\epsfxsize=4in \centerline{ \epsfbox{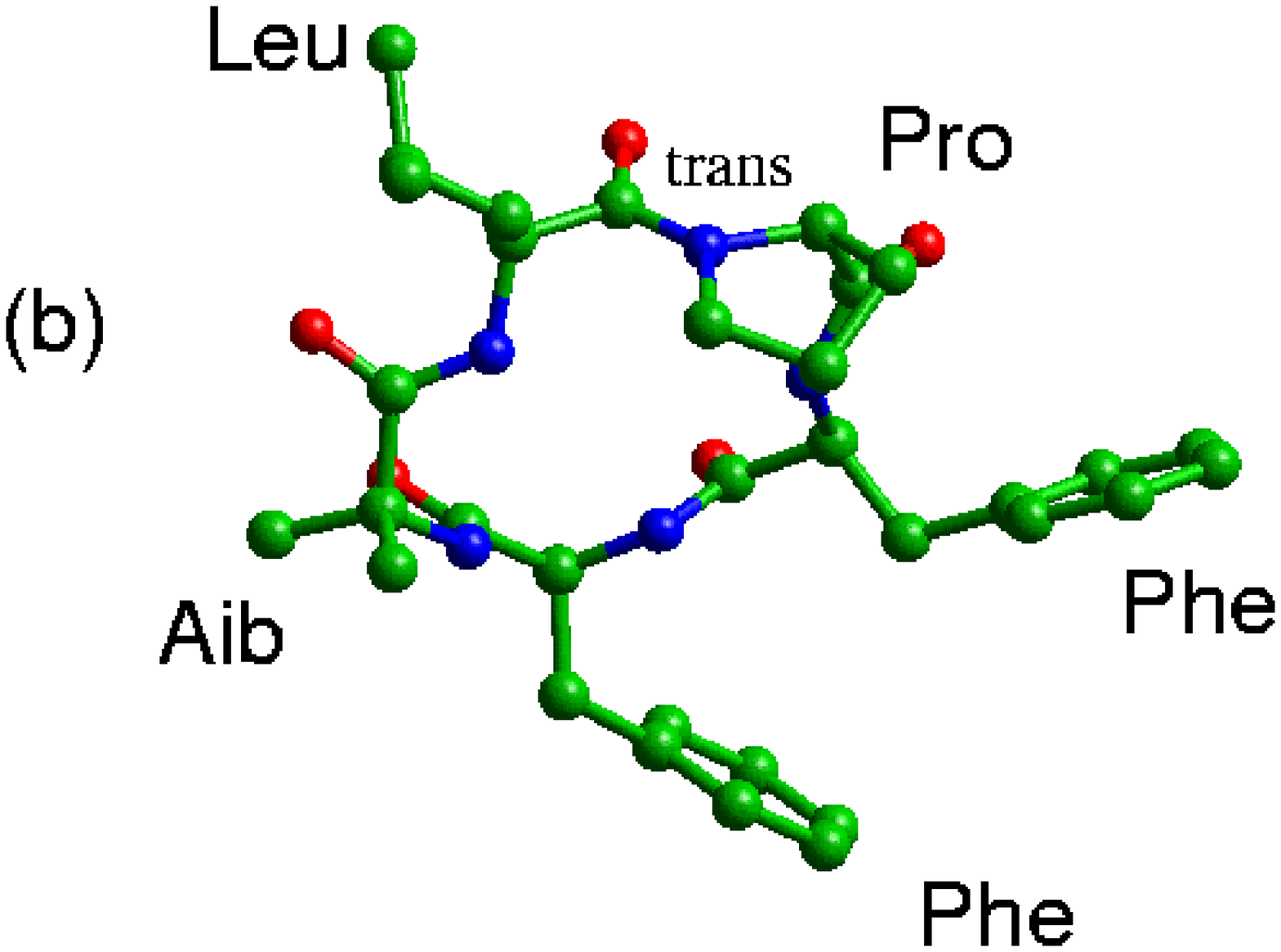}}
Figure \ref{fig:ex2}.
 Wu and Deem, ``Analytical Rebridging Monte Carlo\ldots.''

\newpage
\epsfxsize=5in \centerline{ \epsfbox{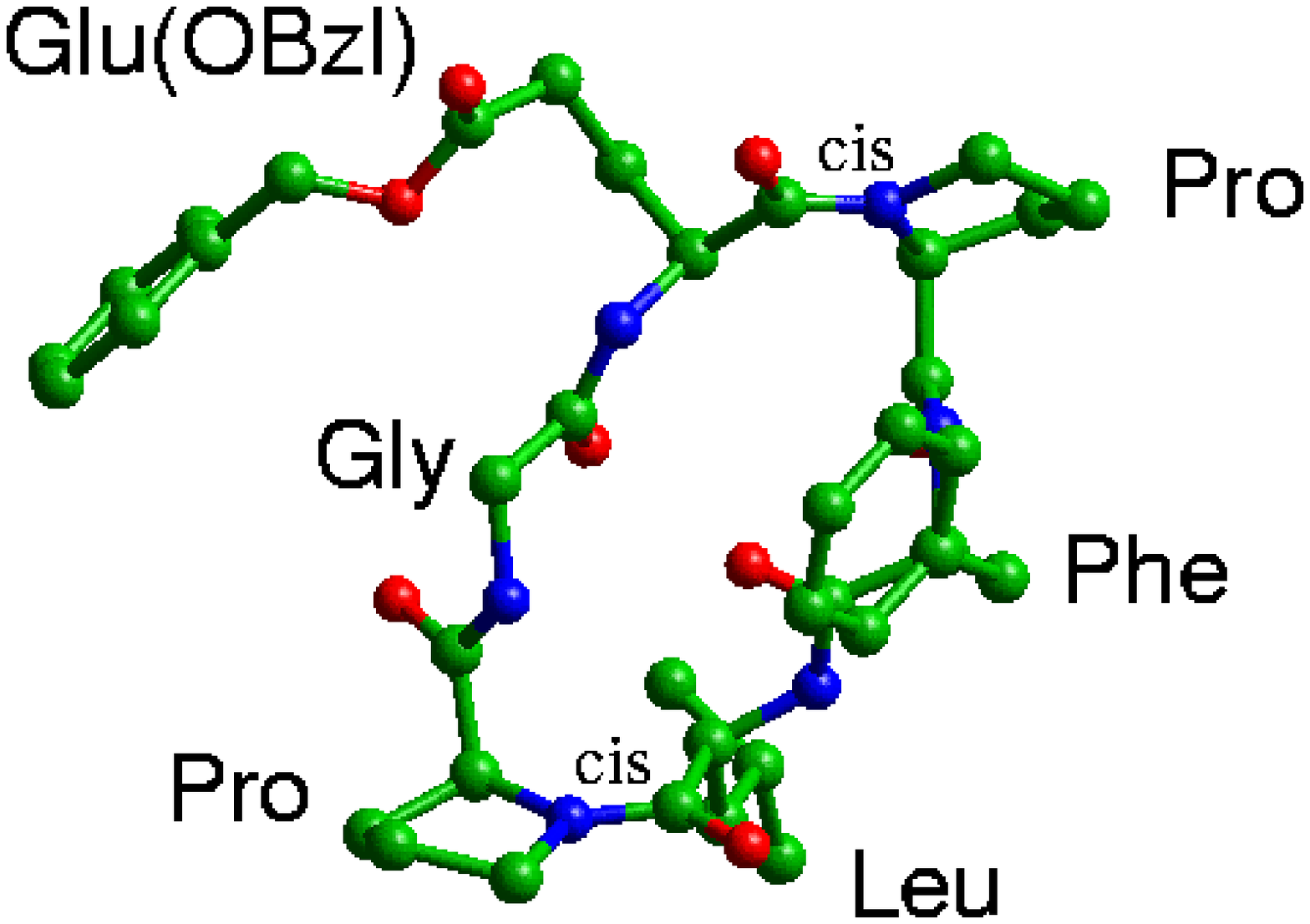}}
Figure \ref{fig:ex4}.
 Wu and Deem, ``Analytical Rebridging Monte Carlo\ldots.''

\newpage
\epsfxsize=5in \centerline{ \epsfbox{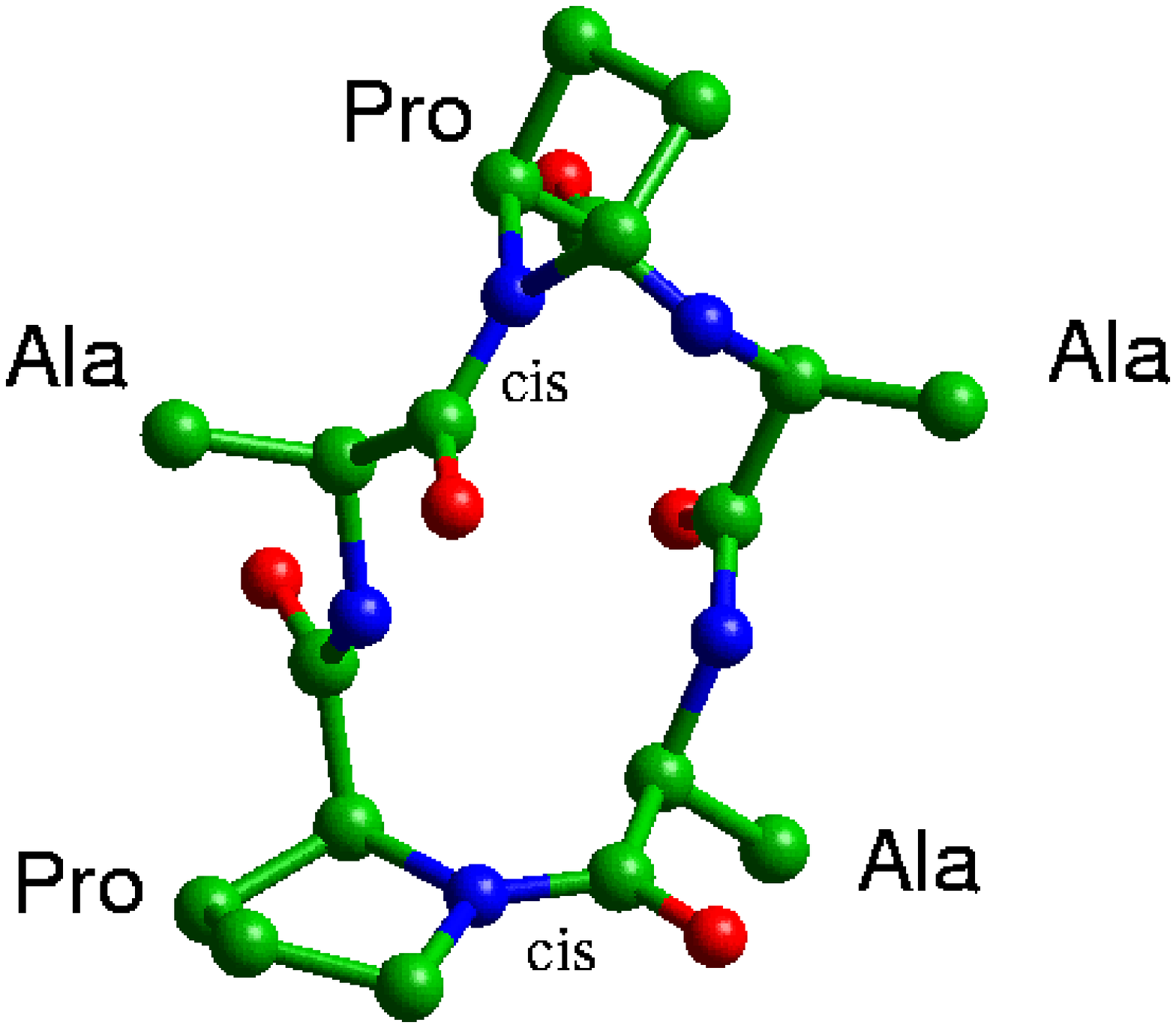}}
Figure \ref{fig:ex7}.
 Wu and Deem, ``Analytical Rebridging Monte Carlo\ldots.''

\newpage
\epsfxsize=5in \centerline{ \epsfbox{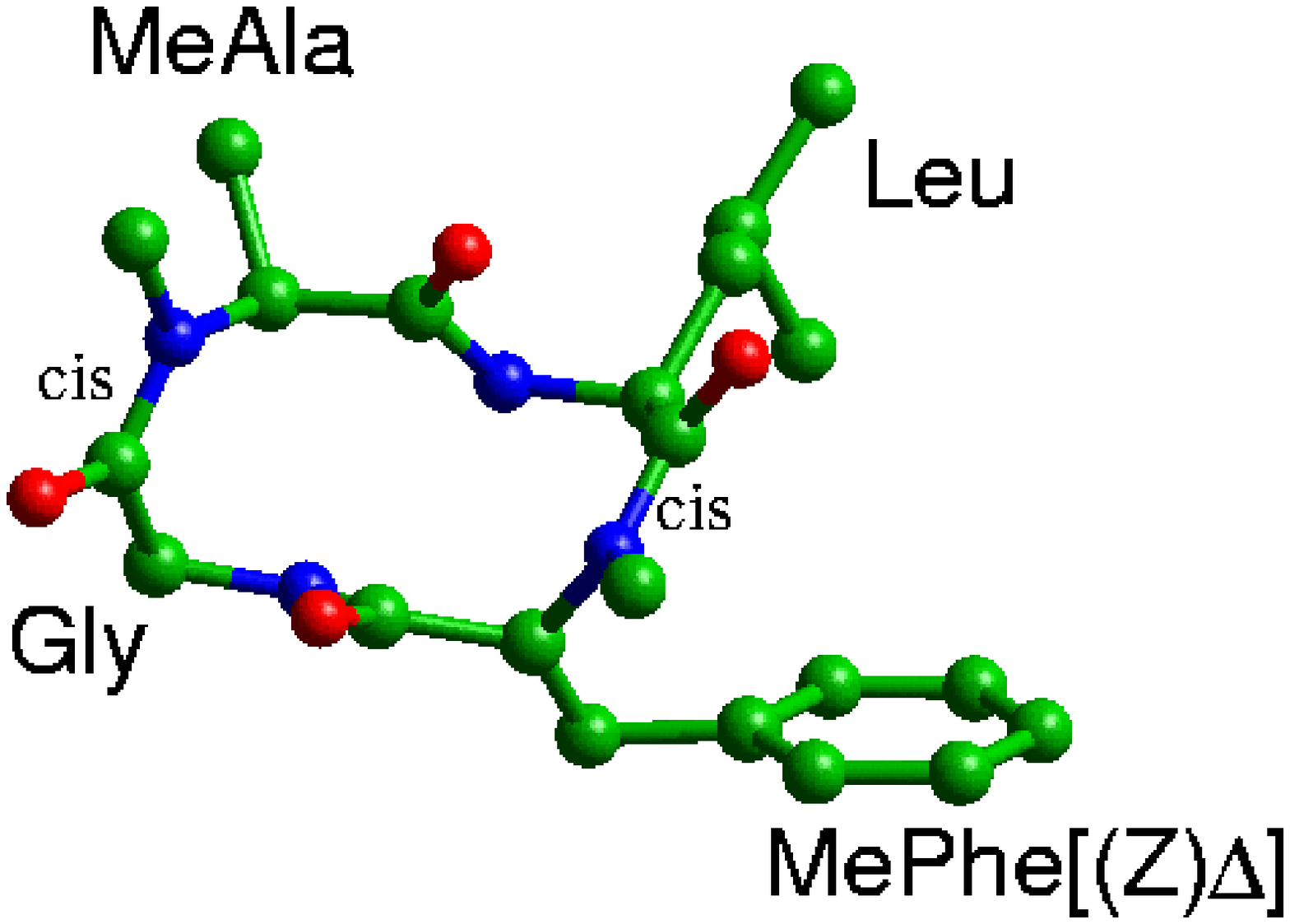}}
Figure \ref{fig:ex8}.
 Wu and Deem, ``Analytical Rebridging Monte Carlo\ldots.''

\newpage
\epsfxsize=5.5in \centerline{ \epsfbox{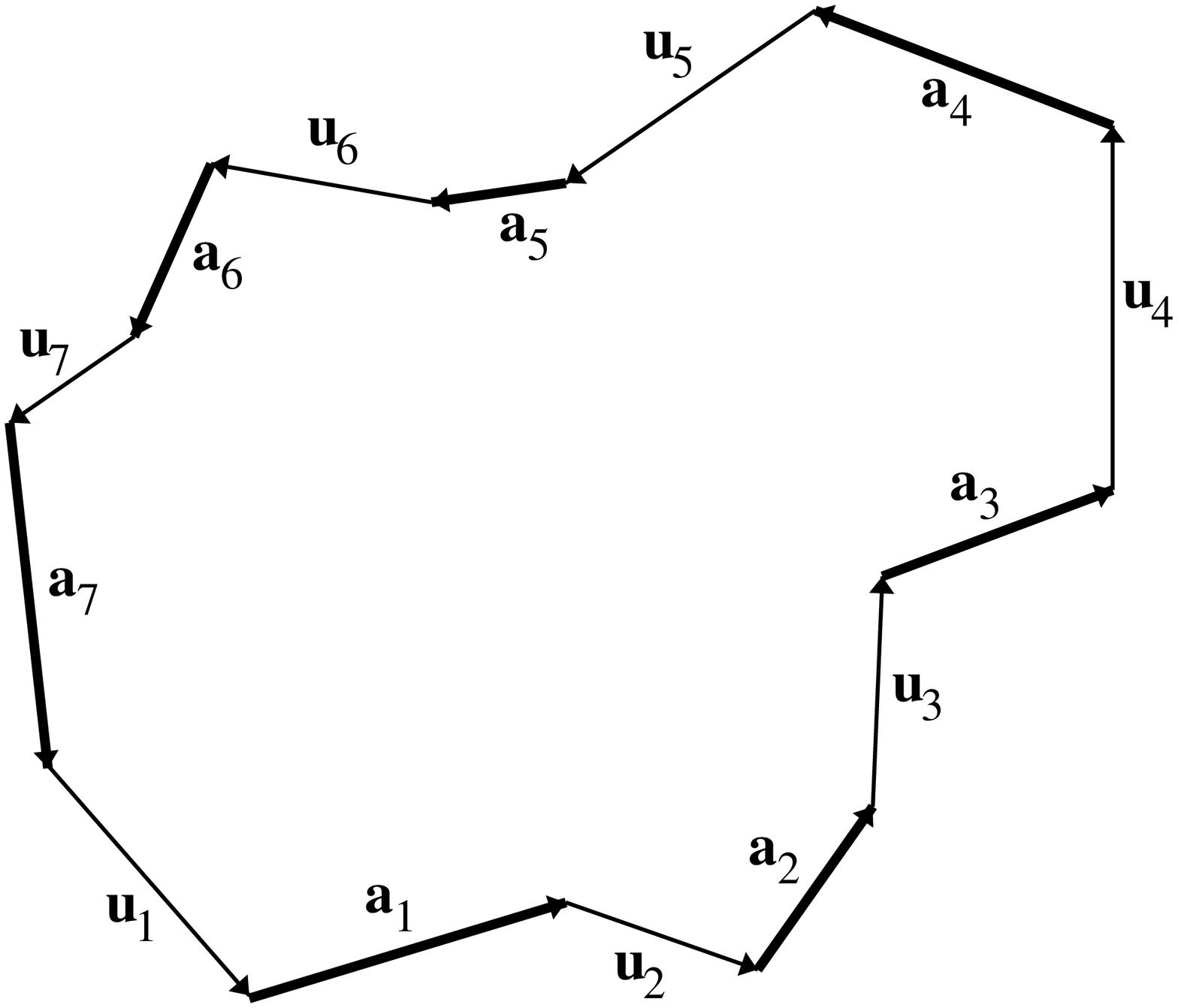}}
\vspace{1.5in}
Figure \ref{fig:7rlink}.
 Wu and Deem, ``Analytical Rebridging Monte Carlo\ldots.''
\end{center}

\end{document}